\documentclass[12pt]{iopart}
\usepackage{graphicx}
\usepackage{aas_macros}

\begin{document}

\title{Electrostatic instability of non-spherical dust in sub-stellar clouds}

\author{C. R. Stark, D. A. Diver and M. I. Swayne}

\address{SUPA, School of Physics and Astronomy, Kelvin Building, University of Glasgow, Glasgow, G12 8QQ, Scotland, UK}
\ead{craig.stark@glasgow.ac.uk}
\vspace{10pt}

\begin{abstract}
 Charged dust clouds play an important role in the evolution of sub-stellar atmospheres through electrical discharges such as lightning events or inter-grain discharges. The consequent plasma activation presents an alternative source of disequilibrium chemistry, potentially triggering a set of chemical reactions otherwise energetically unavailable. 
The aim of this paper is to address the problem of the electrostatic stability of charged spheroidal dust grains in sub-stellar clouds and its impact on inter-grain electrostatic discharges, the available area for atmospheric gas-phase surface chemistry, the particle eccentricity distribution function and observed polarization signatures.
This paper has derived the criterion for the allowed values of dust eccentricity that are electrostatically stable as a function of grain size $a\in[0.2,1.8]~\mu$m, floating potential $\phi_{f}\in[1, 10]$~V and tensile strength $\Sigma_{s}=10^{3}$~Pa. As a consequence of electrostatic instability we also calculate the expected electric field enhancement at the spheroidal poles, the increased surface area of a dust grain, the truncation of the particle eccentricity distribution function and the resultant degree of polarization.
 Dust grains with an eccentricity below a critical value will be electrostatically stable; whereas, grains with an eccentricity above a critical value will be unstable. For example, for a 1~$\mu$m charged dust grain with surface potential of 1 V, the maximum allowable eccentricity is 0.987. As a consequence, electric field strength enhancement at the pole is limited to a factor of $3.5$; the surface area for surface chemistry and the synthesis of chemical products is increased by a factor of 1.45; and the degree of polarization is reduced by a factor of 0.65. 
The results presented here are applicable not only to spheroidal dust grains but any non-spherical dust grains where non-uniform surface electric fields or inhomogeneous tensile strengths could be susceptible to electrostatic instability. In this context electrostatic erosion presents a mechanism that may produce bumpy, irregularly shaped or porous grains.
\end{abstract}

%
%
%
%
%

\section{Introduction}

Charged dust clouds could play a significant role in the dusty atmospheres of brown dwarfs and exoplanets. Triboelectric charging of dust grains \cite{2011ApJ...737...38H} can result in a population of electrically active dust grains leading to large-scale electrical discharges \cite{2013ApJ...767..136H,bailey2013} or smaller-scale inter-grain sparking \cite{2011ApJ...727....4H}. Such ionization events initiate a plasma, injecting a population of free electrons, ions, excited species, radicals and metastables into the local environment. This potentially lowers the activation energy barrier and enhances reaction rates for particular surface reactions to occur, triggering a set of chemical reactions otherwise energetically unavailable. Even a small and transient perturbation to the ambient conditions, initiating non-equilibrium chemistry, can have considerable effects on the outcome of the system. It has been shown that electrical-activity can affect the formation of certain molecules resulting in the weakening of spectral absorption bands such as the 2.7 $\mu$m water line \cite{sorahana2014} and the increase in the abundance of small carbohydrate molecules like CH and CH$_{2}$ \cite{bailey2013}. Charged dust grains can also energise ambient ions in the local environment, accelerating them to the surface of the grain, potentially allowing them to surmount the activation barrier for certain chemical reactions such as the formation of formaldehyde, ammonia, hydrogen cyanide and ultimately glycine \cite{stark2013b}. Electrical activation presents an alternative source of disequilibrium chemistry in atmospheres alongside photochemistry \cite{moses2014} or vertical mixing \cite{saumon2003}.  In contrast to complex plasmas, where there is a sufficient density of charged grains that constitute another plasma component, in this context the emphasis is on transient plasmas (from atmopsheric electrical activity) that have an effect on the ambient grains but those grains do not otherwise contribute to the collective behaviour of the plasma.

Charged dust grains can also impact grain evolution. Negatively charged dust grains can grow electrostatically through the capture of plasma ions from the surrounding ionized medium \cite{2018A&A...611A..91S} and can occur on much faster timescales in comparison to contemporary dust growth via gas-phase surface chemistry. The surface electric field of the charged dust determines how the incoming ions are deposited on the surface and the ultimate resultant geometry of the dust grain \cite{stark2006}. Moreover, \cite{stark2020} investigated the growth of spheroidal dust grains in substellar atmospheres as the result of non-uniform surface electric fields and found that a population of aligned, spheroidal dust grains can produce degrees of polarization $P\approx O(10^{-2} - 1\%)$ consistent with polarimetric observations.

There is a limit to the quantity of charge that can reside on a dust grain that is electrostatically stable and this can have implications for charged dust driven processes such as electrical discharges, dust growth and grain surface chemistry. If the electrostatic stress experienced by the dust grain, as a result of the accumulated charge, exceeds the mechanical tensile stress, the grain will fracture and break apart  \cite{opik1956,rhee1976,fom1992,ros1992,gron2009}.  Stark et al \cite{stark2015} investigated this process in the context of substellar atmospheres and found that the critical grain radius permitted varies from $10^{-9}$ to $10^{-6}$~m depending on the grains' tensile strength for plasma electron temperatures of 10 eV. As a consequence, the local dust clouds become optically thin in the wavelength range $0.1-10$~$\mu$m, with a characteristic peak in the dust optical depth that shifts to higher wavelengths as sub-micrometer particles are destroyed. For charged, non-spherical dust grains the effect of electrostatic instability must consider the non-uniform electric field distribution around the grain e.g., for spheroidal dust grains see \cite{hill81}. An electrostatically unstable dust grain will seek stability by shedding electrons to reduce the electrostatic stress, alter its geometry via erosion, or fracture and break apart.

The purpose of this paper is to address the problem of electrostatic instability of charged non-spherical dust grains in sub-stellar clouds and its impact on inter-grain electrostatic discharges, the available area for atmospheric gas-phase surface chemistry, the particle eccentricity distribution function and observed polarization signatures. It presents a novel application of the electrostatic disruption of charged elongated dust grains in the sub-stellar context building on the work of \cite{hill81} who quantified the electrostatic disruption of elongated or chain-like dust grains in space plasmas such as cometary tails. The paper is structured as follows: Section~\ref{sec_1} derives the physical conditions whereby electrostatic instability and erosion occurs; Section~\ref{sec_2} quantifies the consequences of the electrostatic instability of charged dust grains discussing inter-grain electrostatic discharges, atmospheric surface chemistry, the particle eccentricity distribution function and the effect on polarization signatures; Section \ref{sec_3} summarises the findings of the paper.

\section{Electrostatic instability of spheroidal dust grains \label{sec_1}}
Consider a dust grain immersed in a uniform electron-ion plasma. For a given thermodynamic temperature, the greater mobility of the electrons relative to the ions results in the grain becoming negatively charged. As a result, an electron-depleted plasma sheath forms around the dust grain and a flow of ions towards the grain is initiated. As the dust grain accrues negative charge the probability of further electron attachment decreases and the probability of ion attachment increases, altering the grain's net charge and its potential. The grain's charge varies until a particle flux equilibrium configuration is reached, where the flux of electrons and ions at the grain surface is balanced, resulting in the grain having a constant net negative charge and it's surface potential equal to the floating potential. At this point, the dust grain is surrounded by a plasma sheath with spatial extent of the order of the plasma Debye length, where the potential of the grain is shielded from the plasma.

There exists a limit to the net charge that a dust grain can accumulate, above which grains are unstable: they must evolve by losing charge or mass. If the electrostatic stress acting on the dust grain, as a result of its net negative charge, exceeds the mechanical tensile strength of the grain, then the grain will be electrostatically disrupted and fracture~\cite{stark2015}. For a given electric surface potential, increasing (decreasing) the radius of a spherical dust grain decreases (increases) the electrostatic stress acting on the dust grain. There exists a critical radius where the electrostatic stress acting on the grain exceeds the mechanical tensile stress of the dust grain, electrostatically disrupting the grain and breaking it apart. For radii below the critical radius, the dust grain is unstable to electrostatic disruption; for radii above the critical radius the dust grain is stable to electrostatic disruption.
\begin{figure}[!t]
\centering
\includegraphics[width=0.65\columnwidth]{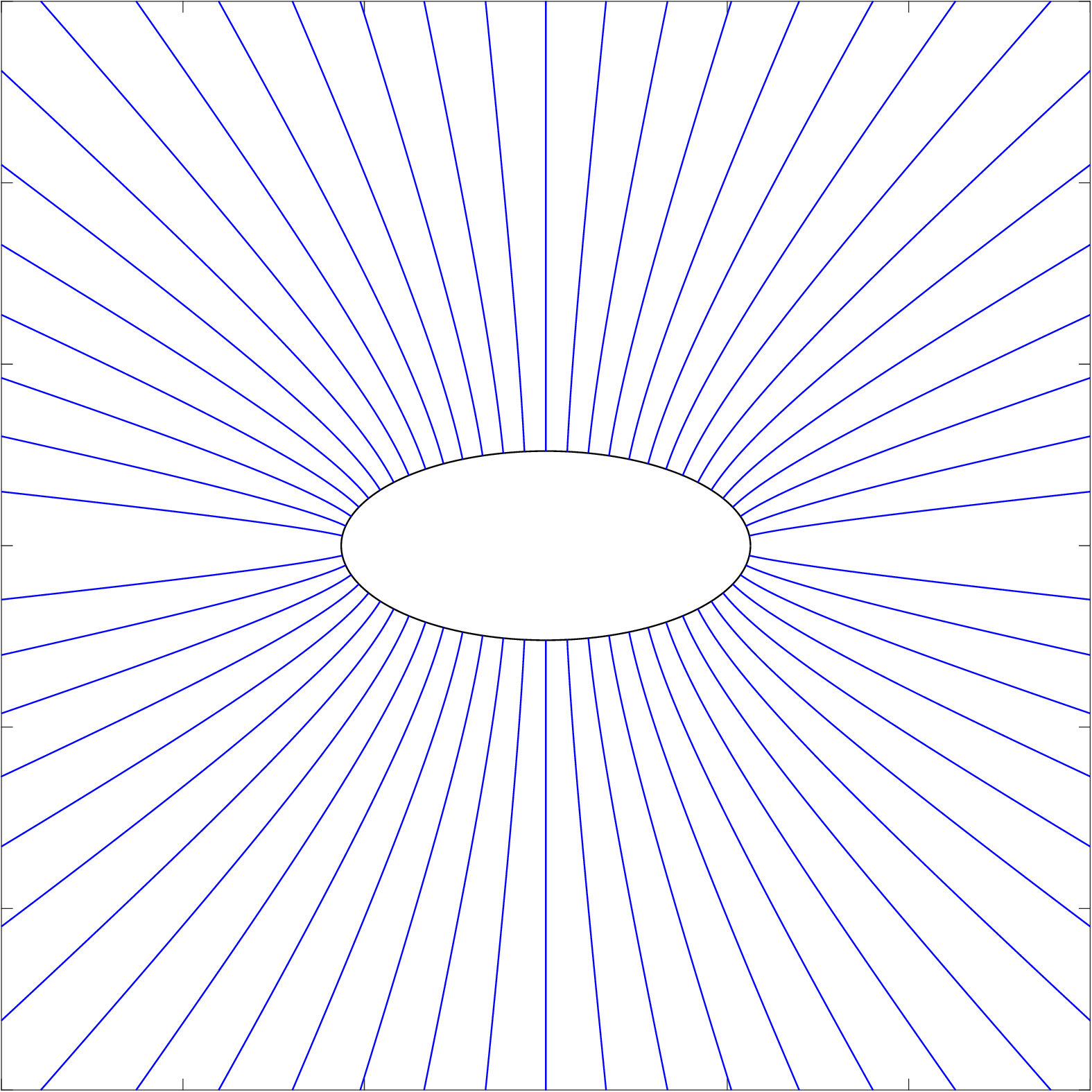}
\caption{Diagram showing the cross-section of a charged spheroidal dust grain (perimeter shown in black) and its associated electric field (blue). The near- and surface electric field is non-uniform as a consequence of the non-uniform distribution of surface charges. }
\label{plot_dust}
\end{figure}

For a spheroidal dust grain the distribution of surface charges is no longer uniform (i.e. non-uniform surface charge density) as in the spherical case and hence the electrostatic stress on the grain is non-uniform (see Figure~\ref{plot_dust}).  As a result, if the local electrostatic stress exceeds the mechanical tensile strength of the dust grain, localised electrostatically-driven erosion can occur altering the geometry of the grain. Consider a conducting prolate spheroidal dust grain.  Due to the concentration of charge at the poles the electrostatic stress is greater than at the equator.  For a grain with uniform tensile strength this means that the grain will mechanically fail (break apart) more readily at the poles, chipping away material and decreasing the eccentricity of the grain until ultimately a spherical grain is produced.

Following~\cite{hill81},~consider a prolate spheroid whose centre is at the origin, with semi-major axis $a$ and minor axes $b$, such that,   
\begin{equation}
\frac{x^{2}}{a^{2}}+\frac{y^{2}+z^{2}}{b^{2}}=1.
\end{equation}
Due to the symmetry of the prolate spheroid, where appropriate the following analysis will be restrict to the $xy$-plane. The electrostatic pressure (or stress) $\xi$ parallel to the $x$ axis (the polar axis) as a function of $x$ can be calculated by considering the spheroid to be divided into two segments by a plane at arbitrary $x$ normal to the polar axis.  Each charge element d$Q$ (located at $(r,\theta,\phi)$ in spherical coordinates) experiences a force in the polar direction due to the charge distribution on the entire spheroid~\cite{hill81}.  The resulting repulsive force $F_{x}$ between the two segments in the $x$ direction is given by
\begin{equation}
F_{x}=\int_{0}^{y^{*}}\frac{\epsilon_{0}E^{2}}{2}2\pi y \textnormal{d}y,
\end{equation}
where
\begin{eqnarray}
E&=&\frac{Q}{4\pi\epsilon_{0}a b^{2}}\left(\frac{x^{2}}{a^{4}}+\frac{y^{2}}{b^{4}}\right)^{-1/2}, \label{ef}\\
y^{*}&=&b(1-x^{2}/a^{2})^{1/2}. \label{ell}
\end{eqnarray}
Equation~\ref{ef} is the electric field of the spheroidal dust grain holding charge $Q$; and equation~\ref{ell} is the equation of an ellipse written in terms of the coordinate $y$.  Evaluating the integral and calculating the electrostatic stress ($\xi(x)=F_{x}/(\pi y^2)$) yields
\begin{equation}
\xi(x)=\frac{Q^{2}\ln{[1+(\lambda^{2}-1)(1-x^{2}/a^{2})]}}{32\pi^{2}\epsilon_{0}b^{4}(\lambda^{2}-1)(1-x^{2}/a^{2})},
\end{equation}
where $\lambda=a/b$.  We are interested in the conditions under which grains of a particular eccentricity can survive, without being disrupted electrostatically and mechanically failing. The conditions for mechanical failure at the grain poles, where the electrostatic stress is the greatest, occur when the polar electrostatic stress exceeds the local mechanical tensile stress, $\Sigma_{s}$.  To probe this, we calculate the polar electrostatic stress by evaluating $\xi(x)$ in the limit of $x\rightarrow a$,
\begin{eqnarray}
\lim_{x \to a}\xi(x)=\frac{Q^{2}}{32\pi^{2}\epsilon_{0}b^{4}}\geq\Sigma_{s},
\end{eqnarray}
therefore the condition for electrostatic disruption at the poles is
\begin{equation}
Q\geq\pi(32\epsilon_{0}\Sigma_{s})^{1/2}a^{2}(1-e^{2}). \label{qcrit}
\end{equation}
Note that as $e\rightarrow 0$ the expression reduces to that of the spherical case~\cite{stark2015}. This expression is applicable to dust charged in a plasma via the collection of electrons and ions or in a gas via triboelectric charging.

The charge $Q$ on a spheroidal grain of semi-major axis $a$ and eccentricity $e=\sqrt{1-b^{2}/a^{2}}$, immersed in a plasma can be related to the plasma floating potential,
\begin{equation}
|\phi_{f}|=\frac{Q}{8\pi\epsilon_{0}ae}\ln{\left(\frac{1+e}{1-e}\right)}. \label{float}
\end{equation}
Combining Equation (\ref{qcrit}) and (\ref{float}), the stability criterion for allowed values of eccentricity, $e$, for a given grain size $a$ and floating potential $\phi_{f}$ can be obtained:

\begin{equation}
\frac{(1-e^{2})}{e}\ln{\left(\frac{1+e}{1-e}\right)}-\delta\leq0, \label{esol}
\end{equation}
where
\begin{equation}
\delta=\left(\frac{2\epsilon_{0}}{\Sigma_{s}}\right)^{1/2}\frac{\phi_{f}}{a}.
\end{equation}
To understand the general behaviour consider a spherical dust grain of radius, $b$, and constant electric surface potential, $\phi_{f}$. Increasing (decreasing) the radius decreases (increases) the magnitude of the surface electric field and hence the electrostatic stress. If the electrostatic stress exceeds the mechanical tensile stress, the dust grain will be electrostatically disrupted and break apart. Increasing (decreasing) the surface electric potential for a given grain size increases (decreases) the electrostatic stress acting on the grain making it less (more) resistant to disruption. Increasing (decreasing) the tensile strength of the dust grain strengthens (weakens) the mechanical tolerance of the dust grain making it more (less) resistant to disruption. 

If the eccentricity of the dust grain increases from zero by allowing one axis to increase in length whilst keeping the remaining axes constant, a spheroid is created. In this scenario, for a constant electric surface potential, the electric field, and hence the electrostatic stress, at the poles of the spheroid increases. If the polar electrostatic stress exceeds the local mechanical tensile stress, the dust grain will become electrostatically unstable and will evolve either by shedding electrons (to reduce the electrostatic stress) or be electrostatically disrupted, preferentially eroding material at the poles of the spheroid. Increasing (decreasing) the overall scale size of the spheroid, while preserving its eccentricity, diminishes (enhances) the electric field strength at the poles rendering the dust grain more (less) stable to electrostatic instability. Hence, spheroidal dust grains with a larger scale size can have a greater eccentricity and still be electrostatically stable; whereas, the eccentricity of smaller scale size dust grains is limited.
\begin{figure}[!t]
\centering
\includegraphics[width=0.75\columnwidth]{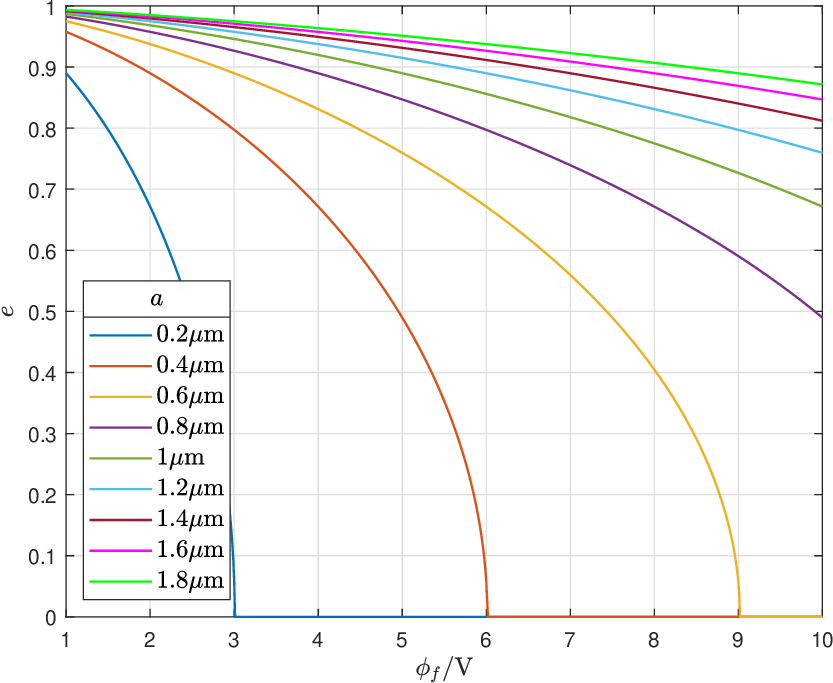}
\caption{Solutions of Equation~(\ref{esol}) as a function of floating potential $\phi_{f}\in[1, 10]$~V and semi-major axis $a\in[0.2,1.8]~\mu$m, numerically calculated using the Newton-Raphson method. The semi-major axis range covers the R, I, J and H near infrared bands of interest with respect to polarimetric observations. Solutions are obtained for a tensile strength $\Sigma_{s}=10^{3}$~Pa. The individual curves represent the critical values of eccentricity, $e_{\rm crit}$, below which dust grains are electrostatically stable ($e<e_{\rm crit}$); above which dust grains are electrostatically unstable ($e>e_{\rm crit}$).}
\label{plot_2}
\end{figure}

Figure~\ref{plot_2} shows solutions of Equation~(\ref{esol}) as a function of floating potential $\phi_{f}\in[1, 10]$~V and semi-major axis $a\in[0.2,1.8]~\mu$m, numerically calculated using the Newton-Raphson method. The semi-major axis range covers the R, I, J and H near infrared bands of interest with respect to polarimetric observations of substellar objects. The solutions were obtained for a tensile strength of $\Sigma_{s}=10^{3}$~Pa, assuming that dust grains in substellar atmospheres are likely to be loose, porous, fluffy aggregates of weakly bonded material. This is consistent with the analysis of dust grains collected from the Jupiter-family comet 67P/Churyumov-Gerasimenko (Schulz et al. 2015). The individual curves represent the critical values of eccentricity, $e_{\rm crit}$, below which dust grains are electrostatically stable ($e<e_{\rm crit}$); above which dust grains are electrostatically unstable ($e>e_{\rm crit}$).  To illustrate the nature of the solutions, in the case where the grain evolves via electrostatic erosion, consider the $a =1\mu$m curve. At $\phi_{f}=1$~V the critical value of eccentricity is approximately $e\approx 0.987$ corresponding to the maximum dust grain eccentricity that is electrostatically stable for a dust grain with constant uniform tensile strength and constant electric surface potential. Hence, the mechanical tensile strength at the poles of the spheroid is sufficient to counteract the electrostatic stress due to the accumulation of charges. As the surface electric potential $\phi_{f}$ increases, the electrostatic stress at the poles increases and exceeds the mechanical tensile strength resulting in the erosion of material. Therefore, the eccentricity of the dust grain decreases until the polar electrostatic stress is again balanced by the mechanical tensile stress.  For smaller dust grain sizes ($a<1\mu$m), at $\phi_{f}=1$~V the critical value of eccentricity is smaller than the $a =1\mu$m case; for example, for $a=0.2\mu$m at $\phi_{f}=1$~V the critical eccentricity is $e\approx0.889$.  For a given electric potential, as the dust grain scale size decreases (increases), the polar electrostatic stress increases (decreases) and the critical eccentricity will decrease (increase) until the stability threshold is reached. In some cases (e.g., $a\in[0.2, 0.6]~\mu$m), as the surface electric potential increases no stable spheroidal equilibrium scenario exists resulting in a spherical dust grain ($e=0$) and, if the electrostatic stress exceeds the mechanical tensile strength of the spherical dust grain then the dust grain will completely fragment.

\section{Consequences of electrostatic instability \label{sec_2}}
\subsection{Inter-grain electrostatic discharge enhancement}
\begin{figure}[!t]
 \begin{center}
\includegraphics[width=0.75\columnwidth]{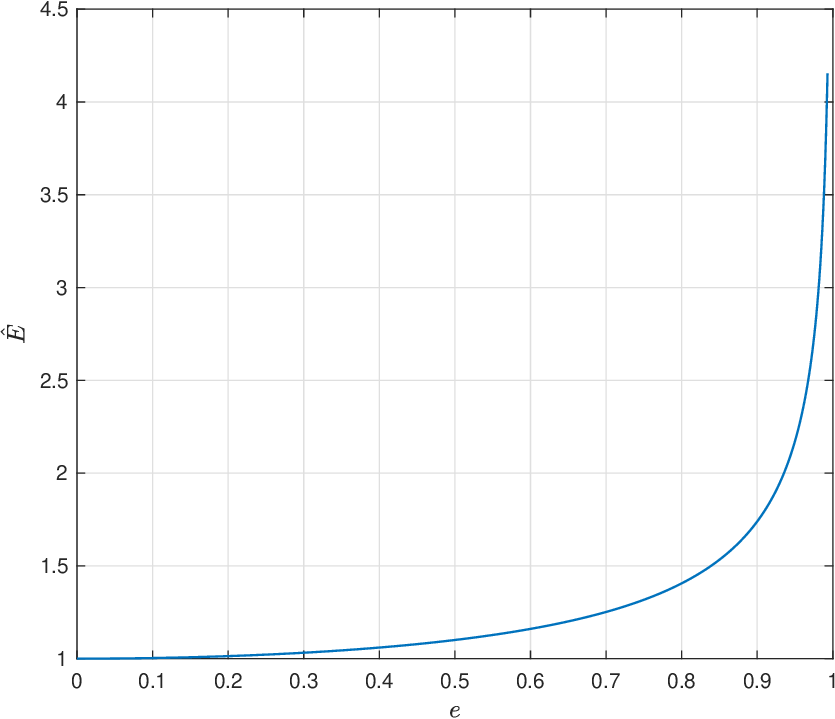}
\caption{Electric field strength enhancement, $\hat{E}$, at the pole of a spheroidal dust grain as a function of grain eccentricity, $e$, in contrast to a spherical dust grain for a given volume. As the eccentricity of a dust grain increases from zero, the electric field strength at the poles increases up to a maximum value.}
\label{plot_3}
 \end{center}
\end{figure}

\begin{figure}[!t]
\centering
\includegraphics[width=0.75\columnwidth]{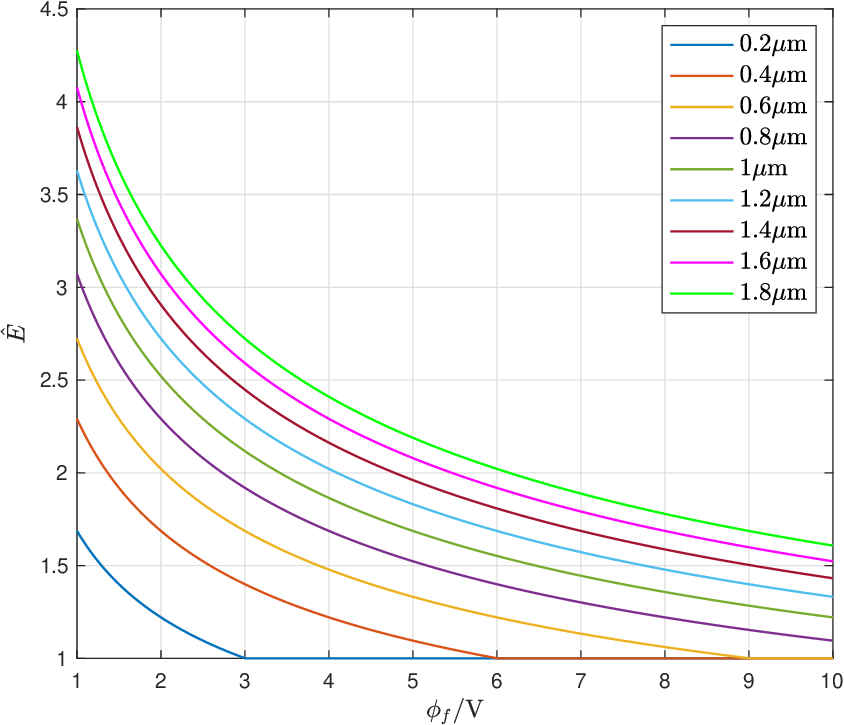}
\caption{Polar electric field strength enhancement, $\hat{E}$, as a function of surface electric potential, $\phi_{f}$, and semi-major axis $a$ (grain scale size). The enhancement is calculated using the electrostatic stability criteria obtained using Equation~(\ref{esol}) and the solutions plotted in Figure~\ref{plot_2}.}
\label{plot_4}
\end{figure}

Sparking between charged dust grains can electrically activate the environment allowing chemical reactions to occur that would not be achievable through thermal excitation alone. Due to the tip-effect of charged objects, spheroidal dust grains will have enhanced electric field strengths at their poles enhancing inter-grain electrical discharges. For a spherical dust grain of radius $R$ and spheroidal dust grain with semi-axes $a$ and $b$ to have the same volume, the following constraint must be satisfied,
\begin{equation}
R=a(1-e^{2})^{1/3}. \label{vol}
\end{equation}
Therefore, the electric field strength on the surface of a spherical dust grain of radius $R$ is
\begin{equation}
E_{0}=\frac{Q}{4\pi\epsilon_{0}R^{2}}=\frac{Q}{4\pi\epsilon_{0}a^{2}(1-e^{2})^{2/3}}.
\end{equation}
For a given volume of dust grain, the electric field strength at the pole of a spheroidal dust grain of semi-axes $a$ and $b$ is
\begin{equation}
E=\frac{Q}{4\pi\epsilon_{0}b^{2}}. \label{E_pole}
\end{equation}
To obtain this result equation~(\ref{ef}) has been evaluated for $(x,y)=(a,0)$. A measure of the enhancement of the electric field strength at the pole of a spheroidal dust grain can be defined as,
\begin{equation}
\hat{E}=\frac{E}{E_{0}}=(1-e^{2})^{-1/3}. \label{E_basic}
\end{equation}
Figure~\ref{plot_3} shows the enhancement of the electric field strength $\hat{E}$ as a function of eccentrcity, $e$.  As the eccentricity of a dust grain increases from zero, the electric field strength at the poles increases up to a maximum value of $\hat{E}\approx4.2$ corresponding to an arbitrary maximum eccentricity of $e\approx0.9897$.  Please note that the parameter $\hat{E}$ preserves the volume of the dust grain and so larger values of eccentricity will always yield optimum field enhancements. Figure~\ref{plot_4} demonstrates the polar electric field strength enhancement calculated using the electrostatic stability criteria calculated using Equation~(\ref{esol}) and the solutions plotted in Figure~\ref{plot_2}. The individual curves represent the critical electric field strength enhancement, $\hat{E}_{\rm crit}$, corresponding to the critical eccentricities as shown in Figure~\ref{plot_2}. In general, small scale size grains with large surface electric potentials and high eccentricities will yield the highest electric field enhancements. However, the effect of electrostatic erosion is to preferentially remove this population from the dust particle eccentricity distribution. Therefore, for a given surface electric potential, dust grains of a smaller scale size will have a lower electric field enhancement in comparison to larger scale sizes since the latter are able to retain stability for higher values of eccentricity. As the surface electric potential increases in magnitude, electrostatic erosion removes the dust grains with high values of eccentricity that would otherwise yield strong electric field enhancement. 

 Helling et al \cite{2013ApJ...767..136H} investigated the electrostatic breakdown characteristics, including the critical electric field strengths, to determine the conditions under which substellar dust clouds undergo electrical discharge events. The minimum electric field strength to achieve breakdown is $E_{\rm t, min}=Bp$, where $B$ is a constant that is dependent upon the composition of the atmospheric gas and $p$ is the atmospheric gas pressure. Therefore, for a given grain volume, a spheroidal dust grain can achieve the minimum electric field strength $E_{\rm t, min}$ for atmospheric pressures a factor $\hat{E}$ larger than can be achieved by a spherical dust grain.

\begin{figure}[!t]
\centering
\includegraphics[width=0.85\columnwidth]{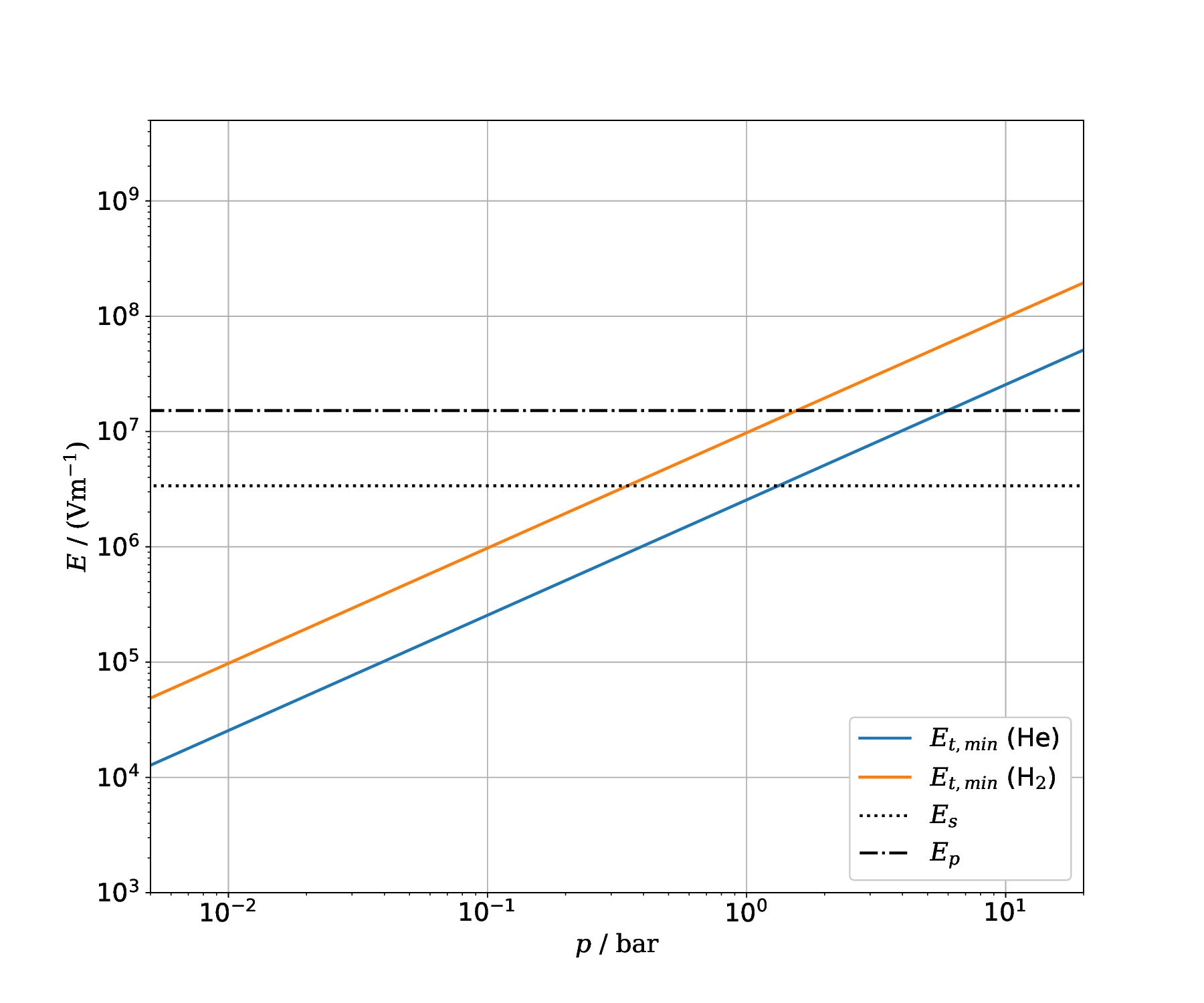}
\caption{The minimum electric field strength to achieve breakdown, $E_{\rm t, min}=Bp$, as a function of atmospheric pressure, $p$ for H$_{2}$ and He, where $B=97.5$~V/(m Pa) and $B=25.5$~V/(m Pa) repsectively~\cite{2013ApJ...767..136H}. The plot also shows the electric field strength at the pole of a spheroidal grain, $E_{p}$, and of a spherical grain, $E_{s}$, for a given volume and $\phi_{f}= 1$~V, $a=1\mu$m and $e=0.987$.}
\label{plot_4b}
\end{figure}

To illustrate the effect of the electric field enhancement on the conditions for electrical breakdown, let's consider the electric field strength at the pole of a spheroidal grain, $E_{p}$, and for a spherical grain, $E_{s}$, for a given volume,
\begin{eqnarray}
E_{p} &=& \frac{2\phi_{f}}{a}\frac{e}{(1-e^2)}\left[\ln{\left(\frac{1+e}{1-e}\right)}\right]^{-1}, \label{E_p}\\
E_{s} &=& \frac{\phi_{f}}{a(1-e^2)^{1/3}}. \label{E_s}
\end{eqnarray}
Equation (\ref{E_p}) is obtained from equation (\ref{E_pole}) using $b^2=a^2(1-e^2)$ and eliminating $Q$ via equation (\ref{float}); and, equation (\ref{E_s}) is obtained using $E_{s}=\phi_{f}/R$ and equation (\ref{vol}). Figure \ref{plot_4b} shows the minimum breakdown field $E_{\rm t, min}$ for He and H$_{2}$ as a function of atmospheric pressure (see~\cite{2013ApJ...767..136H}); and $E_{p}$ and $E_{s}$ for $\phi_{f}= 1$~V, $a=1\mu$m and $e=0.987$.  Due to the enhanced electric field at the poles the dust grain is more likely to meet the electrical breakdown conditions in cloudy substellar atmospheres. 
\subsection{Impact on the available area for atmospheric gas-phase surface chemistry}
\begin{figure}[!t]
\centering
\includegraphics[width=0.75\columnwidth]{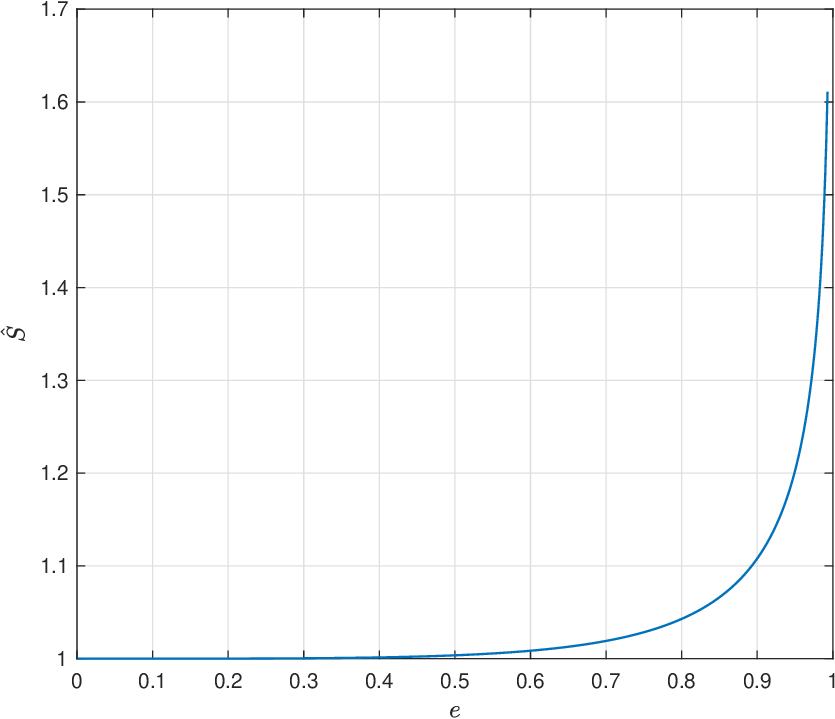}
\caption{Surface area enhancement, $\hat{S}$, presented by a spheroidal dust grain as a function of grain eccentricity $e$ in contrast to a spherical grain for a given volume. As the eccentricity of a dust grain increases from zero, its surface area increases up to a maximum value.}
\label{plot_5}
\end{figure}

\begin{figure}[!t]
\centering
\includegraphics[width=0.75\columnwidth]{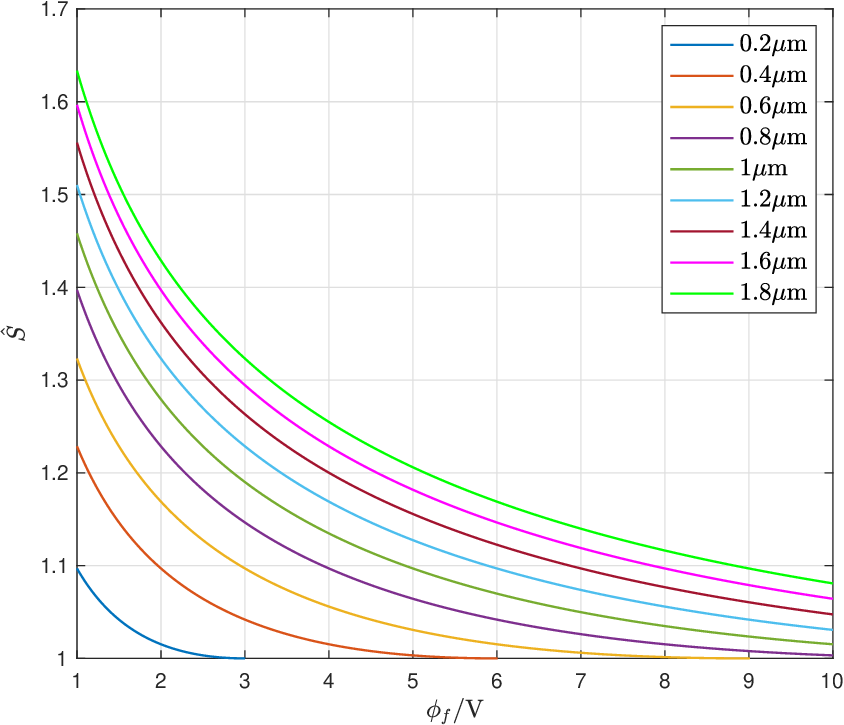}
\caption{Surface area enhancement, $\hat{S}$, as a function of surface electric potential, $\phi_{f}$, and semi-major axis $a$ (grain scale size). The enhancement is calculated using the electrostatic stability criteria obtained using Equation~(\ref{esol}) and the solutions plotted in Figure~\ref{plot_2}.}
\label{plot_6}
\end{figure}

For a given volume, the surface area of a spheroid is greater than that of a sphere and so the atmospheric gas sees a greater effective surface area for surface chemistry. The surface area of the spherical dust grain is
\begin{equation}
S_{0}=4\pi R^{2}=4\pi a^{2}(1-e^{2})^{2/3},
\end{equation}
and the surface area of the spheroid is
\begin{equation}
S=2\pi b^{2}\left(1+\frac{\arcsin{(e)}}{e\sqrt{1-e^{2}}}\right).
\end{equation}
In the limit of $e\rightarrow0$, $S\rightarrow4\pi b^{2}$. A measure of the increased surface area presented by a spheroidal dust grain in contrast to a spherical grain for a given volume can be defined as,
\begin{equation}
\hat{S}=\frac{S}{S_{0}}=\frac{1}{2}(1-e^{2})^{1/3}\left(1+\frac{\arcsin{(e)}}{e\sqrt{1-e^{2}}} \right)
\end{equation}
Figure~\ref{plot_5} shows the increased surface area presented by a spheroidal dust grain in contrast to a spherical grain for a given volume. As the eccentricity of a dust grain increases from zero, its surface area enhancement increases up to a maximum value of $\hat{S}\approx1.6$ corresponding to an arbitrary maximum eccentricity of $e\approx0.9897$. Figure~\ref{plot_6} demonstrates the surface area enhancement calculated using the electrostatic stability criteria obtained using Equation~(\ref{esol}) and plotted in Figure~\ref{plot_2}. The individual curves represent the critical surface area enhancement, $\hat{S}_{\rm crit}$, corresponding to the critical eccentricities plotted in Figure~\ref{plot_2}, but for dust grains of a larger scale size will always yield greater surface areas. For a given scale size and volume, increasing the eccentricity will yield greater surface areas. As a result of electrostatic erosion, for a given scale size, greater eccentricities occur at lower values of surface electric potential $\phi_{f}$ (Figure~\ref{plot_2}), and so as $\phi_f$ decreases (increases) the surface area enhancement increases (decreases).

To see the impact on atmospheric chemistry and plasma-surface interactions, without loss of generality, consider the simple surface reaction $A+B\rightarrow AB$, where $A$ is a reactant initially in the ambient gas-plasma that impinges on a dust surface and interacts with a molecule $B$ residing on the surface, resulting in a chemical reaction to produce the chemical product $AB$. The rate equation for such an interaction can be expressed as $\dot{n}_{AB}=kn_{A}n_{B}$, where $n_{B}$ ($n_{AB}$) is the areal concentration of species $B$ ($AB$) on the dust surface; $n_{A}$ is the volume concentration of species $A$ in the ambient gas-plasma; and, $k$ is the rate constant for the reaction. Assuming $n_{A}$ and $n_{B}$ are constant and $S_{0}$ is the surface area of a spherical dust grain, the change in the number of molecules $AB$ on the surface of the dust grain in a time interval $\Delta t$ as a consequence of the chemical reaction, can be expressed as $\Delta N_{AB}=S_{0}\Delta n_{AB}=kn_{A}n_{B}S_{0}\Delta t$. Therefore, for a given volume of dust grain the fractional increase in $N_{AB}$ due to the surface area of a spheroidal dust grain, $S$, is 
\[
\frac{kn_{A}n_{B}S\Delta t}{kn_{A}n_{B}S_{0}\Delta t}=\frac{S}{S_{0}}=\hat{S}.
\]
Therefore, the greater the surface area of the dust grain, the more chemical products produced for a given time interval $\Delta t$ and hence the faster the reaction. Additionally, the resulting fragmented material from electrostatic disruption can increase the effective volume and the collective reactive surface area of the reactant leading to faster reactions e.g., fragmenting a 1 $\mu$m dust grain into an equivalent volume of smaller particulates of $10^{-8}$ m size increases the surface-area-to-volume ratio by a factor of $10^{2}$. 

\begin{figure}[!t]
\centering
\includegraphics[width=0.75\columnwidth]{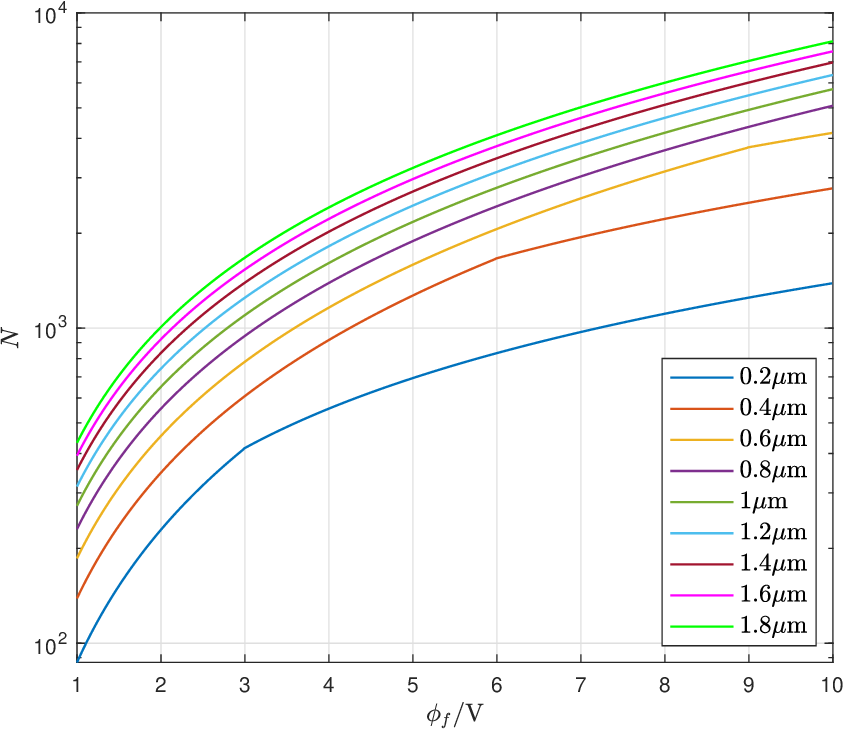}
\caption{Net number of electron charges residing on the surface of a dust grain, $N$, as a function of surface electric potential, $\phi_{f}$, and semi-major axis $a$ (grain scale size). The net charge is calculated using the electrostatic stability criteria obtained using Equation~(\ref{esol}) and the solutions plotted in Figure~\ref{plot_2}.}
\label{plot_9}
\end{figure}

Furthermore, the net charge residing on the surface of a dust grain will also enhance surface chemistry: gas reagents arriving at the surface will encounter the electrons and might be excited enough by the local electric field to get over the reaction barrier threshold; or the incoming molecules pick up electrons from the surface by attachment. Figure~\ref{plot_9} shows the net number of electron charges residing on the surface of a dust grain obtained using the electrostatic stability criteria plotted in Figure~\ref{plot_2}. The individual curves show the net number of electrons residing on the grain surface, obtained from Equation (\ref{float}), given by,
\[
N = \frac{8\pi \epsilon_{0}ae|\phi_{f}|}{q_{e}\ln{[(1+e)/(1-e)]}},
\]
where $q_{e}$ is the charge of an electron. In the case where $e=0$, then $N=4\pi\epsilon_{0}a |\phi_{f}|/q_{e}$. 
\subsection{Truncation of the particle eccentricity distribution function}
\begin{figure}[!t]
\centering
\includegraphics[width=0.75\columnwidth]{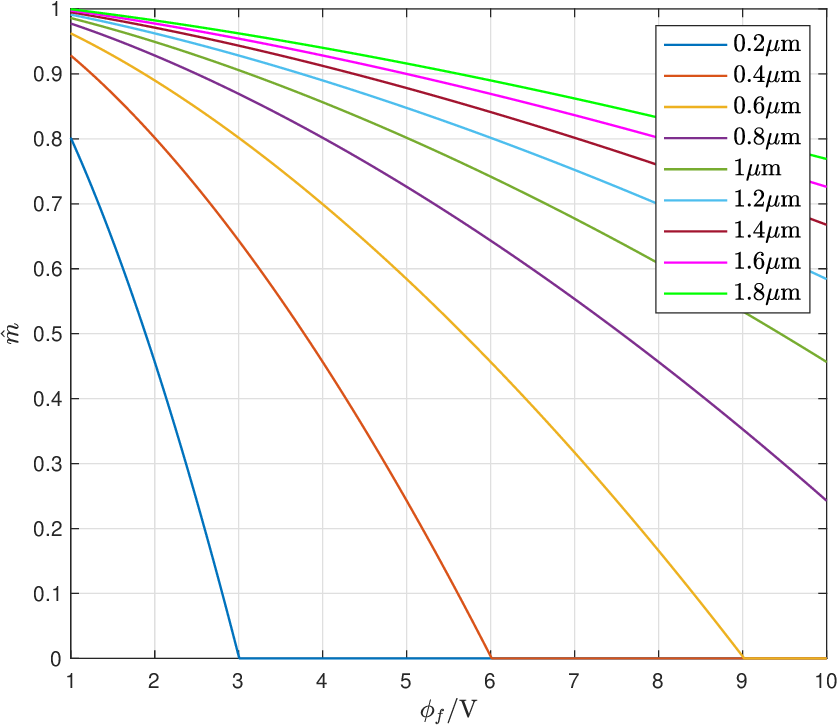}
\caption{Modification of the dust particle eccentricity distribution function, $\hat{m}$, as a function of surface electric potential, $\phi_{f}$, and semi-major axis $a$ (grain scale size). The modification parameter, $\hat{m}$, is calculated using the electrostatic stability criteria obtained using Equation~(\ref{esol}) and the solutions plotted in Figure~\ref{plot_2}.}
\label{plot_7}
\end{figure}

The particle eccentricity distribution function will be truncated as a result of the electrostatic erosion of spheroidal dust grains. To quantify this effect, following  \cite{stark2020}, consider a continuous flat distribution for a given value of semi-major axis $a$,
\begin{equation}
f(m) = \frac{1}{m_{\rm max}},~~~~~\textnormal{for}~~~m\in[0,m_{\rm max}],
\end{equation}
with mean value $\langle m\rangle_{0}=\frac{1}{2}m_{\rm max}$; where $m=e^{2}$; and, $m_{\rm max}=e_{\rm max}^{2}$ is the maximum value of eccentricity squared. A measure of the modification of the particle eccentricity function can be defined as,
\begin{equation}
\hat{m}=\frac{\langle m\rangle}{\langle m\rangle_{0}}=\frac{m_{\rm crit}}{m_{\rm max}},
\end{equation}
where $\langle m\rangle=\frac{1}{2}m_{\rm crit}$; and, $m_{\rm crit}=e_{\rm crit}^{2}$ is the threshold value of eccentricity squared, above which spheroidal dust grains are unstable to electrostatic erosion; below which they are stable. $m_{\rm crit}$ is a function of the length scale of the dust grain characterised by the semi-major axis, $a$. In absence of electrostatic erosion, the parameter $\hat{m}=1$; in the presence of electrostatic erosion $\hat{m}<1$. Figure~\ref{plot_7} shows the modification of the particle eccentricity distribution size expressed in terms of $\hat{m}$ obtained using Equation~(\ref{esol}) and plotted in Figure~\ref{plot_2}. For a given grain scale size, increasing the surface electric potential, increases the electrostatic stress at the poles. Dust grains with an eccentricity greater than the critical threshold for electrostatic erosion are unstable and are removed from the particle eccentricity distribution function. Dust grains with an eccentricity lower than the critical threshold are stable to electrostatic erosion and remain part of the distribution function. As a result, the mean eccentricity derived from the particle eccentricity distribution function decreases in value in contrast to the absence of electrostatic erosion. As the grain scale size decreases, the dust grains become increasingly unstable to electrostatic erosion. Therefore, for a continuous flat eccentricity distribution function, the critical threshold for electrostatic erosion occurs for progressively smaller values of eccentricity, truncating the distribution function more severely. As a result, the mean eccentricity decreases in value in contrast to the scenario where there is no electrostatic erosion.
\subsection{Effect on polarization signature}

\begin{figure}[!t]
\centering
\includegraphics[width=0.75\columnwidth]{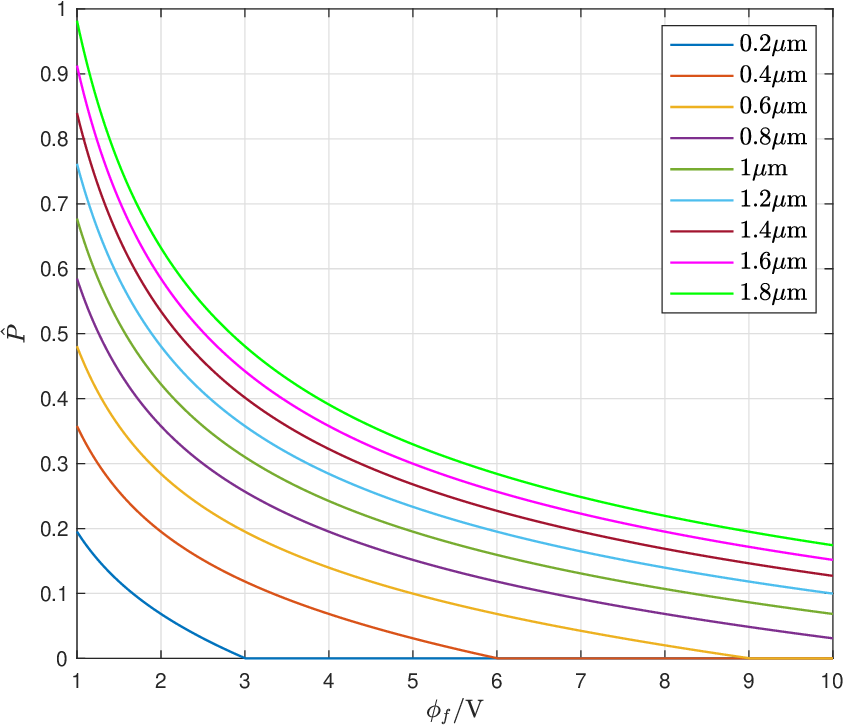}
\caption{Effect on the polarization signature, $\hat{P}$, as a function of surface electric potential, $\phi_{f}$, and semi-major axis $a$ (grain scale size), calculated using the electrostatic stability criteria obtained using Equation~(\ref{esol}) and the solutions plotted in Figure~\ref{plot_2}.}
\label{plot_8}
\end{figure}

Following \cite{stark2020}, an order of magnitude estimation of the expected degree of polarization, $P$, as a consequence of a population of aligned spheroidal dust grains can be expressed as
\begin{equation}
P = 50.01\zeta L\kappa_{d}n_{d}m_{d}m(1-m)^{-1/2}
\end{equation}
where $\kappa_{d}$ is the dust opacity; $m_{d}=\frac{4}{3}\pi a^{3}(1-m)\rho_{m}$ is the dust grain mass; $\rho_{m}$ is the mass
density of a dust grain; $n_{d}$ is the dust particle number density; $L$ is a characteristic atmospheric length scale reflecting the spatial extent of an atmospheric region containing a population of aligned spheroidal dust grains; and, $\zeta\in[0,1]$ is a free parameter that quantifies that not all grains will be perfectly aligned, nor will the entire atmospheric region be populated by spheroidal dust grains. A measure of the effect of electrostatic erosion on the polarization of a population of dust grains can be defined as,
\begin{equation}
\hat{P}=\frac{P}{P_{0}}=\frac{m_{\rm crit}}{m_{\rm max}}\left(\frac{1-m_{\rm crit}}{1-m_{\rm max}}\right)^{-1/2},
\end{equation}
where,
\begin{equation}
P_{0} = 50.01\zeta L\kappa_{d}n_{d}m_{d}m_{\rm max}(1-m_{\rm max})^{-1/2},
\end{equation}
is the expected degree of polarization from spheroidal dust grains that obtain a maximal eccentricity and
\begin{equation}
P =  50.01\zeta L\kappa_{d}n_{d}m_{d}m_{\rm crit}(1-m_{\rm crit})^{-1/2},
\end{equation}
is the expected degree of polarization from spheroidal dust grains whose maximal eccentricity is truncated by electrostatic erosion. Please note that $\hat{P}$ is valid for a given dust grain scale size. Figure~\ref{plot_8} shows the effect on the degree of polarization expressed in terms of $\hat{P}$ obtained using Equation~(\ref{esol}) and plotted in Figure~\ref{plot_2}. For a given scale size, as the surface electric potential increases in magnitude, dust grains are not permitted to grow to the maximal eccentricity due to their instability to electrostatic erosion. As a consequence, the degree of polarization decreases.  The dust grain scale size is a proxy for the wavelength of the resulting polarization signature, and at smaller wavelengths the strength of polarization is diminished. This is a result of smaller sized dust grains being more susceptible to electrostatic erosion. Inspite of the limiting effect on the degree of polarization from electrostatic erosion, values of polarization consistent with observable values are still achievable. For example, consider the solution corresponding to $a\approx0.8\mu$m (I-band) and $\phi_{f}=2$~V, with $\hat{P}\approx0.358$. For typical values, $n_{d}\approx10^{10}$\,m$^{-3}$ (e.g. \cite{helling2014}), $\kappa_{d}\approx10^{-4}$\,m$^{2}$kg$^{-1}$ (e.g. \cite{helling2008b,lee2016}), $\zeta\approx10^{-1}$ and $L\approx10^{7}$\,m (e.g.~\cite{isabel2018}), the degree of polarization, $P = P_{0}\hat{P}$, is of the order $\approx10^{-3}\%$. Similarly, for $a\approx1.22\mu$m ($a\approx0.6\mu$m)  corresponding to the J-band (R-band) and $\phi_{f}=2$~V, with $\hat{P}\approx0.481$ ($\hat{P}\approx0.284$), the degree of polarization is of the order $\approx10^{-2}\%$ ($\approx10^{-3}\%$). These are first approximation calculations, please note that the degree of polarization is sensitive to variations in the implicated parameters.
\section{Discussion \label{sec_3}}
This paper has addressed the problem of electrostatic stability of charged non-spherical dust grains in sub-stellar clouds and its impact on inter-grain electrostatic discharges, atmospheric surface chemistry, the particle eccentricity distribution function and observed polarization signatures. It has built on the work of \cite{hill81} who quantified the electrostatic stability of elongated dust grains in space plasmas such as cometary tails. We have derived the criterion for allowed values of eccentricity that a dust grain can have as a function of grain size $a\in[0.2,1.8]~\mu$m , floating potential $\phi_{f}\in[1, 10]$~V and tensile strength $\Sigma_{s}=10^{3}$~Pa. Dust grains with an eccentricity below a critical value will be electrostatically unstable; whereas, grains with an eccentricity above a critical value will be stable. For example, for a 1~$\mu$m charged dust grain with surface potential of 1 V, the maximum allowable eccentricity is 0.987. As a consequence, a 1~$\mu$m a charged dust grain of eccentricity 0.987 can enhance the electric field strength at the pole of a spheroidal dust grain by a factor of $3.5$; it can increase the surface area for surface chemistry and the synthesis of chemical products by a factor of 1.45; and reduce the degree of polarization by a factor of 0.65. The results presented here are applicable not only to spheroidal dust grains but any non-spherical dust grains where non-uniform surface electric fields or inhomogeneous tensile strengths could yield electrostatic erosion. In this context electrostatic erosion presents a mechanism that may produce bumpy, irregularly shaped or porous grains.

The impact of electrostatic erosion discussed may have observational consequences. Consider an atmosphere, populated by spheroidal dust grains that triboelectrically charge and participate in inter-grain sparking. As a result, the local environment will become populated with free electrons and ions increasing the fractional ionization. When a sufficient fractional ionization is reached the partially ionised medium will contribute towards the charging of the dust grains via the collection of ions and electrons, permitting greater surface electric potentials, triggering further inter-grain sparking and increasing the plasma density and the electron temperature. In addition, the created plasma will contribute towards the growth of the dust grains via ion accretion~\cite{stark2006} and facilitate the growth of further spheroidal dust grains with a range of eccentricities~\cite{stark2020}. As a consequence, the effective surface area available for surface chemistry leads to enhanced chemical products being produced on the grain, depleting the atmospheric gas. However, if consequentially the dust becomes unstable to electrostatic erosion, the particle eccentricity distribution function of the spheroidal dust grains will be truncated, affecting the efficiency of inter-grain sparking and the maintenance of a threshold fractional ionization to facilitate the enhancement of inter-grain sparking. As a result, the plasma density will significantly decrease until the population of spheroidal dust grains recovers and the cycle resumes. Such a cyclic process would produce periodic sparking events with a characteristic period dependent upon the timescale to grow spheroidal dust grains of sufficient eccentricity. For example, spheroidal dust grains grown via plasma deposition \cite{2018A&A...611A..91S}, at an atmospheric pressure of $1$~bar, will have a growth rate in the range $\gamma\in[10^{-9}-10^{-1}]$~ms$^{-1}$, which for grain scale sizes of $1~\mu$m corresponds to a growth characteristic timescale of $t\in[10^{-5}, 10^{3}]$~seconds~\cite{stark2020}.  Please note that the characteristic timescale is sensitive to the local plasma conditions. This suggests that correlated variability in radio (including associated optical if present) from the discharge events and near-infrared polarization emission, from the population of spheroidal dust grains, with period $10^{3}$ seconds could be attributable to such a scenario. Furthermore, the characteristic timescale is also consistent with the lightning statistics from \cite{gabi2016}, where the number of lightning flashes per transit for HD189733b could be in the range $10^{5}-10^{6}$ which corresponds to a flash period of $10^{-2}-10^{-1}$ seconds.

Although discussed within the context of substellar atmospheres, the work presented is also applicable to other areas where charged non-spherical dust is implicated such as in the Earth's atmosphere \cite{mallios2021}, volcanic dust clouds  \cite{buckland2021} and lunar dust \cite{richard2011}. For example, optical polarimetry observations of a Saharan dust episode indicated the presence of vertically aligned, non-spherical dust particles in the Earth's atmopshere as the result of the collective organization of charged dust cloud particles under the influence of the self-electric field permeating the cloud \cite{2007ACP.....7.6161U}. Furthermore, volcanic ash plumes can have a population of elongated shaped dust particles \cite{shoji2018} that can contribute towards volcanic lightning \cite{cimarelli2014} and permits the possibility of diagnosing further properties of the volcanic ash cloud via polarimetry \cite{miffre2011}. The results presented here contribute to the understanding of charged non-sperhoidal dust particles and their observational manifestion.

%
%
%
%
\ack
CRS and MIS are grateful for funding from STFC via grant number ST/X000885/2. 
\\
\bibliographystyle{iopart-num}
\bibliography{spheroid_pap}
\end{document}